\documentclass[conference]{IEEEtran}
\IEEEoverridecommandlockouts
\usepackage{cite}
\usepackage{amsmath}
\usepackage{amsfonts}
\usepackage{amssymb}
\usepackage{amsthm}
\usepackage{tikz}
\usepackage{cases}
\usepackage{graphicx}
    \graphicspath{{../}}
    \DeclareGraphicsExtensions{.pdf}
\usepackage[caption=true,font=footnotesize]{subfig}
\usepackage{multirow}
\usepackage{booktabs}
\usepackage{url}
\usepackage{xtab}
\usepackage{array}
\usepackage{algorithm}
\usepackage{algorithmic}
\usepackage{enumerate}
\usetikzlibrary{arrows}
\theoremstyle{plain}

\newtheorem{theorem}{Theorem}

\theoremstyle{definition}

\begin{document}

\title{Building Capacity-Achieving PIR Schemes with Optimal Sub-Packetization over Small Fields}
\author{\IEEEauthorblockN{Jingke Xu,~Zhifang Zhang}\\
\IEEEauthorblockA{\fontsize{9.8}{12}\selectfont KLMM, Academy of Mathematics and Systems Science, Chinese Academy of Sciences, Beijing 100190, China\\
School of Mathematical Sciences, University of Chinese Academy of Sciences, Beijing 100049, China\\
Emails:  xujingke14@mails.ucas.edu.cn, zfz@amss.ac.cn}
}
\maketitle
\thispagestyle{empty}
\begin{abstract}
Consider $N$ servers with replicated databases containing $M$ records. Suppose a user wants to privately retrieve one record by accessing the servers such that the identity of the retrieved record is secret against any up to $T$ servers. A scheme designed for this purpose is called a $T$-private information retrieval ($T$-PIR) scheme. Three indexes are concerned for PIR schemes:
\begin{itemize}
  \item[(1)] rate, indicating the amount of retrieved information per unit of downloaded data. The highest achievable rate is characterized by the capacity;
  \item[(2)] sub-packetization, reflecting the implementation complexity for linear schemes;
  \item[(3)] field size. We consider linear schemes over a finite field.
\end{itemize}
In this paper, a general $T$-PIR scheme simultaneously attaining the optimality of almost all of the three indexes is presented. Specifically, we design a linear capacity-achieving $T$-PIR scheme with sub-packetization $\!dn^{M-1}\!$ over a finite field $\mathbb{F}_q$, $q\geq N$. The sub-packetization $\!dn^{M-1}\!$, where $\!d\!=\!{\rm gcd}(N,T)\!$ and $\!n\!=\!N/d$, has been proved to be optimal in our previous work. The field size
is reduced by an exponential factor in our scheme comparing with existing capacity-achieving $T$-PIR schemes.
\end{abstract}

\section{Introduction}\label{sec1}
The problem of private information retrieval (PIR) involves a database containing $M$  records stored across $N$ servers and a user who wants to privately retrieve one record by accessing the servers. The privacy requirement means any colluding subset of at most $T$ servers knows nothing about the identity of the retrieved record. Since it is closely related to cryptography \cite{BIKO12ACCC:SharePIR} and coding theory \cite{Yekhanin07PHD:LDC&PIR}, PIR has become a central research topic in the computer science literature since it was first introduced by Chor et al. \cite{CKGS95FOCS:PIR} in 1995. A survey on PIR can be found in \cite{Gas04EATCS:SurveyonPIR}.

A central issue in PIR is minimizing the communication cost which is  measured by the total number of bits transferred from the user to the servers (i.e. the query size) and those from  servers to the user (i.e. the answer size). In the initial setting of PIR where each record is one bit long, the least communication cost achieved by now is $\!M^{O(\frac{1}{\log\log M})}\!$ \cite{Dvir&Gopi15STOC:2PIR,Efremenko09:LDC}. Recently, an information-theoretic reformulation of the PIR problem \cite{CHY15ISIT:PIR} allows all records to be arbitrarily large, which is more relevant to real-life applications. In this sense, the query size is always negligible compared to the answer size. Thus communication cost can be measured by taking only the answer size into account. Specifically, define the {\it rate} of a PIR scheme as the ratio between the size of the retrieved record and the answer size, and define the {\it capacity} as the supremum of the rate over all PIR schemes. Obviously, the capacity characterizes the highest rate that can be achieved for a given PIR problem.

Sun and Jafar \cite{Sun&Jafar16:CapacityPIR} first proved that the capacity of PIR for replicated non-colluding servers (i.e., $\!T\!=\!1$) is {\small$\frac{1-1/N}{1-(1/N)^{M}}$}. They further proved in \cite{Sun&Jafar16:ColludPIR} the capacity for the colluding case (i.e., $T$-PIR, $1\!\leq\!T\!<\!N$) is {\small$\frac{1-T/N}{1-(T/N)^{M}}$} and designed a capacity-achieving PIR scheme over a finite filed $\mathbb{F}_q$, where $\!q\!\geq\!\max\{N^2T^{\!M\!-\!2},N^2(N\!-\!T)^{\!M\!-\!2}\}$. Meantime, the capacity for PIR with symmetric privacy was determined in \cite{Sun&Jafar16:CapaSymmPIR}, and that for PIR with MDS coded non-colluding servers was studied in \cite{Bana&Uluk16:CapacityPIRCoded,Wang&Sko16:CapaSymmePIR}. In this work, we focus on PIR problems from replicated servers and only non-symmetric privacy is concerned. A detailed model for the PIR problem studied in this work is presented in Section II.

A key step for determining the capacity is to design a general capacity-achieving PIR scheme. All existing schemes are implemented by dividing each record into segments (taken as elements in the finite field $\mathbb{F}_q$) and querying from each server some combinations of the segments. We call the number of segments contained in each record that are necessary for the scheme implementation as {\it sub-packetization} which reflects implementation complexity for linear schemes where only linear operations over the segments are involved. The schemes provided in \cite{Sun&Jafar16:CapacityPIR,Sun&Jafar16:ColludPIR} both have sub-packetization $N^M\!$.
In \cite{Sun&Jafar16:OptimalPIR} the sub-packetization was firstly reduced to $\!N^{M\!-\!1}$ for replicated servers and $\!T\!=\!1$. For general values of $T$, we recently proved in \cite{Zhang&Xu17:OptimalSubpacketization} that the optimal sub-packetization for linear capacity-achieving $T$-PIR schemes is $dn^{M-1}$, where $d={\rm gcd}(N,T),n=N/d$.

Since all existing PIR schemes are linear over a finite field, the field size is also an important metric related to the computational complexity. A recent improvement on the field size for capacity-achieving $T$-PIR schemes is obtained in \cite{Zhang&Xu17:OptimalSubpacketization} which reduces the field size  by a factor of $Nd^{M-2}$ comparing with the scheme in \cite{Sun&Jafar16:CapacityPIR}, that is, the scheme in \cite{Zhang&Xu17:OptimalSubpacketization} operates over a finite field $\mathbb{F}_q, q\!\geq\! \max\{Nt^{\!M\!-\!2},N\!(n\!-\!t)^{\!M\!-\!2}\}$, where $\!t=\!T/d$. However, the field size is still larger than $Nt^{M-2}$ which expands exponentially as $M$ grows.

The main contribution of this work is a great reduction on the field size for $T$-PIR schemes which simultaneously achieve the capacity and the optimal sub-packetization. Specifically, we design a linear capacity-achieving $T$-PIR scheme with sub-packetization $\!dn^{M\!-\!1}\!$ over a finite field $\mathbb{F}_q, q\!\geq\!N$. Comparing with previous schemes, the field size is reduced by an exponential factor and become a constant for a fixed storage system.

The rest of this paper is organized as follows. First, the PIR model is formally defined in
Section \ref{sec2}. Then an example of the $T$-PIR scheme is presented in Section III to explain our design idea. The general descriptions of our scheme are given in Section IV. Finally, Section V concludes the paper.

\section{Notations and the PIR model}\label{sec2}
For an integer $n\!\in\!\mathbb{N}$, we denote by $[n]$ the set $\{1,...,n\}$. For a vector $Q=(q_1,...,q_n)$ and a subset $\Gamma=\{i_1,...,i_m\}\subseteq [n]$, denote $Q_\Gamma=(q_{i_1},...,q_{i_m})$. Most vectors in this paper are row vectors and they are denoted by the capital letters (eg. $W,Q$).

Denote the $M$ records as {\small$W_1,...,W_M$} and the $N$ servers as {\small$\!\rm{Serv}^{(1)}\!,...,\rm{Serv}^{(N)}\!$}. In this paper, we focus on PIR for replicated servers, that is, assume each server stores all the $M$ records. Then suppose a user wants to privately retrieve $W_\theta$ for some $\theta\!\in\![M]$. Basically, a PIR scheme consists of two phases:
\begin{itemize}
  \item {\bf Query phase. }Given $\theta\in[M]$ and some random resources $S$, the user computes {\small${\rm Que}(\theta,S)=(Q_\theta^{(1)},...,Q_\theta^{(N)})$}, and sends {\small$Q_\theta^{(j)}$} to {\small$\rm{Serv}^{(j)}$} for $1\!\leq\! j\!\leq\!N$. Note that $S$ and $\theta$ are private information only known by the user, and {\small${\rm Que}(\cdot,\cdot)$} is the {\it query function} defined by the scheme.
  \item {\bf Response phase. }For $1\!\leq\!j\!\leq\!N$, the $j$th server {\small$\rm{Serv}^{(j)}$}  at receiving {\small$Q_\theta^{(j)}$}, computes {\small${\rm Ans}^{(j)}(Q_\theta^{(j)},W_{[M]})=A_\theta^{(j)}$} and sends it back to the user, where {\small${\rm Ans}^{(j)}(\cdot,\cdot)$} is {\small$\rm{Serv}^{(j)}$}'s {\it answer function} defined by the scheme.
\end{itemize}
To design a PIR scheme is to design the functions ${\rm Que}$ and ${\rm Ans}^{(j)}\!, 1\!\leq\!j\!\leq\!N$, such that the following two conditions are satisfied:
\begin{itemize}
\item[(1)]{\it Correctness: }The user can recover {\small$W_\theta$} after collecting all answers from the $N$ servers, i.e., {\small$H(W_\theta|A^{[N]}_\theta\!,Q^{[N]}_\theta\!,S,\theta)\!=\!0$}, where {\small$H(\cdot|\cdot)$} is the conditional entropy.
\item[(2)]{\it Privacy: } For any $\Gamma\!\subseteq\![N]$ with $|\Gamma|\!\leq\!T$, the serves in $\Gamma$ have no information on $\theta$ even if they collude with each other,  i.e., {\small$I(\theta;Q^{\Gamma}_\theta,A^{\Gamma}_\theta,W_{[M]})=0$}, where {\small$I(\cdot~;\cdot)$} denotes the mutual information.

\end{itemize}
Particularly, if {\small$W_i\!\in\!\mathbb{F}_q^L\!$}  and the answer {\small$A_\theta^{(j)}$} turns out to be linear combinations of {\small$W_i$}'s for $\!1\!\leq\!i\!\leq\!M\!$ and $1\!\leq\!j\!\leq\!N$, we call the scheme a {\it linear} $T$-PIR scheme, and call $L$ the {\it sub-packetization} of the scheme.

Define the {\it rate} $\mathcal{R}$ of a PIR scheme by
{\small$$\mathcal{R}=\frac{H(W_\theta)}{\sum_{i=1}^{N}H(A^{(i)}_\theta)}\;,$$}
that is, $\mathcal{R}$ characterizes the amount of retrieved information per unit of downloaded data. Furthermore, the supremum of the rate over all PIR schemes that work for a given PIR problem is called the {\it capacity}. It was determined in \cite{Sun&Jafar16:ColludPIR} that the capacity of $T$-PIR schemes is {\small$\frac{1-T/N}{1-(T/N)^{M}}$}.

The next theorem determines the optimal sub-packetization for linear capacity-achieving $T$-PIR schemes.
\begin{theorem} \cite{Zhang&Xu17:OptimalSubpacketization}\label{thm2} Suppose $M\!\geq\!2, N\!>\!T\!\geq1$. For all linear capacity-achieving $T$-PIR schemes with $M$ records and $N$ replicated servers, its sub-packetization must be no less than $dn^{M-1}$, where $d={\rm gcd}(N,T),n=N/d$. Moreover, the lower bound can be obtained generally.
\end{theorem}

\section{An Example for $M=3,N=3,T=2$}
Suppose $M\!=\!3,N\!=\!3,T\!=\!2$. Since our $T$-PIR scheme needs sub-packetization $L\!=\!dn^{M\!-\!1}\!=\!9$, each record is regarded as a $9$-dimensional row vector over $\mathbb{F}_q$, i.e., {\small$W_1,W_2,W_3\!\in\!\mathbb{F}_q^9$}, where $q\geq 3$. WLOG, assume the user wants to retrieve $W_1$, i.e., $\!\theta\!=\!1$. The scheme is described in two parts:

{\it Part I: Mixing-Rearranging.}

Let $S_1,S_2,S_3$ be three random matrices chosen by the user independently and uniformly from all $9\times9$ invertible matrices over $\mathbb{F}_q$. And, $S_1,S_2,S_3$ are only known by the user.  Define

\vspace{-0.4cm}

{\small\begin{align} \label{eqb}\renewcommand*{\arraystretch}{0.85}
   &(a_{ij})_{3\times 3}={\rm Mat}_{3\times3}(W_1S_1) \notag\\
   &(b_{ij})_{3\times 3}={\rm Mat}_{3\times3}(W_2S_2[:,(1\!:\!6)](I_3\otimes G) )\notag\\
   &(c_{ij})_{3\times 3}={\rm Mat}_{3\times3}(W_3S_3[:,(1\!:\!6)](I_3\otimes G))
\end{align}}\\

\vspace{-0.8cm}

\noindent where  {\small$\!{\rm Mat}_{s\times t}(\cdot)\!$} is to write out a $s\times t$ matrix row by row from a $st$-dimensional vector. For example, {\small${\rm Mat}_{2\times3}\big((1,2,0,0,1,2)\big)\!=\!\begin{pmatrix}
1&2&0\\0&1&2
\end{pmatrix}$}. And,
{\small$S_i[:,(1\!:\!6)]$} denotes the matrix composed of the first $6$ columns of {\small$S_i$}, $\otimes$ denotes the  Kronecker product, {\small$I_3$} is the $3\times 3$ identity matrix, and $G$ is a $2\times 3$ generator matrix of some fixed $[3,2]$ MDS code over $\mathbb{F}_q$.

Specifically, write the three matrices generated in (\ref{eqb}) as

\vspace{-0.3cm}
{\small\begin{equation}\label{eq22}{\scriptsize\begin{pmatrix}
a_{11}&a_{12}&a_{13}\\a_{21}&a_{22}&a_{23}\\a_{31}&a_{32}&a_{33}
\end{pmatrix},\begin{pmatrix}
b_{11}&b_{12}&b_{13}\\b_{21}&b_{22}&b_{23}\\a_{31}&b_{32}&b_{33}
\end{pmatrix},\begin{pmatrix}
c_{11}&c_{12}&c_{13}\\c_{21}&c_{22}&c_{23}\\c_{31}&c_{32}&c_{33}
\end{pmatrix}},\end{equation}}\\
\vspace{-0.6cm}

\noindent then one can see each $a_{ij}$, $1\leq i,j\leq 3$, is a random linear combination of the $9$ coordinates of $W_1$, and $(a_{ij})_{3\times3}$ is actually an invertible transformation of $W_1$, while $(b_{ij})_{3\times3}$ is composed of three independent codewords (as rows) in the $[3,2]$ MDS code, encoded from $6$ symbols each of which is a random linear combination of coordinates of $W_2$. The construction of $(c_{ij})_{3\times3}$ is similar to that of $(b_{ij})_{3\times3}$.

{\it Part II: Combining.}

We complete description of the scheme by listing in Table \ref{fg1} the answers from all servers.

\begin{table}[ht]
\centering
\setlength{\abovecaptionskip}{0.cm}
\setlength{\belowcaptionskip}{-0.2cm}
\begin{tikzpicture}[scale=2]

%\draw[->] (1.79,0.15) arc (55:-55:0.15);
%\draw[->] (1.3,0.15) arc (55:-55:0.15);
%\draw[->] (1.3,-0.23) arc (55:-55:0.15);
%\draw[->] (1.3,-0.52) arc (55:-55:0.15);

\draw[dotted](-1.93,0.08)--(1.93,0.08);
\draw[dotted](-1.93,-0.29)--(1.93,-0.29);
\draw[dotted](-1.93,-0.5)--(1.93,-0.5);

\draw(-1.93,0.7)--(-1.93,-0.7);
\draw(-0.55,0.7)--(-0.55,-0.7);
\draw(0.83,0.7)--(0.83,-0.7);
\draw(1.92,0.7)--(1.92,-0.7);
\node at (2.13,0.25){$\rightarrow$\scriptsize(a1)};
\node at (2.13,-0.15){$\rightarrow$\scriptsize(a2)};
\node at (2.13,-0.4){$\rightarrow$\scriptsize(a1)};
\node at (2.13,-0.6){$\rightarrow$\scriptsize(a2)};
\node at (0,0){\footnotesize\begin{tabular}{c c c}
\specialrule{0.09em}{0pt}{1.5pt}$\rm{Serv}^{(1)}$ & $\rm{Serv}^{(2)}$ & $\rm{Serv}^{(3)}$\\\specialrule{0.09em}{1pt}{1.5pt}
$a_{11},b_{11},c_{11}$&&$a_{13},b_{13},c_{13}$\\
&$a_{22},b_{22},c_{22}$&$a_{23},b_{23},c_{23}$\\\specialrule{0em}{1.6pt}{1.6pt}
 &$a_{12}\!+\!b_{12}, a_{32}\!+\!c_{12}$ & \\
$a_{21}\!+\!b_{21}, a_{31}\!+\!c_{21}$&&\\\specialrule{0em}{1.6pt}{1.6pt}
$b_{31}\!+\!c_{31}$&$b_{32}\!+\!c_{32}$&\\\specialrule{0em}{1.6pt}{1.6pt}
&&$a_{33}\!+\!b_{33}\!+\!c_{33}$\\
\specialrule{0.09em}{0.5pt}{0pt}
\end{tabular}};
\end{tikzpicture}
\caption{\scriptsize Answer table of the $T$-PIR scheme for
$(M\!=\!3,N\!=\!3,T\!=\!2)$ and $\theta\!=\!1$.}
\label{fg1}
\end{table}

It can be seen that the answers are all sums of the symbols $a_{ij},b_{ij},c_{ij}$. Because $a_{ij}$'s are generated from $W_1$, the record desired by the user, we call $a_{ij}$ the {\it desired symbol}. Accordingly, we call $b_{ij}, c_{ij}$ the {\it interference symbol}. Likewise, all sums that do not involve $a_{ij}$'s, such as $b_{ij}+c_{i'j'}$, are  called {\it interference sums}. The sums that involve both $a_{ij}$'s and interference symbols, such as $a_{ij}+b_{i'j'}$, $a_{ij}+c_{i'j'}$, are called {\it mixed sums}. Basically, the answer table is built up by iteratively applying the following two steps:

\begin{itemize}
 \item[(a1)] \emph{Enforce record symmetry within each server.}
 \item[(a2)]\emph{Form mixed sums by combining new desired symbols with recoverable interferences.}
\end{itemize}

As shown in Table \ref{fg1}, the record symmetry within each server has been enforced in the first two lines. That is, one symbol from each record is queried from {\small$\!{\rm Serv}^{(1)}\!$}, and also from {\small$\!{\rm Serv}^{(2)}\!$}, while two symbols from each record are queried from {\small$\!{\rm Serv}^{(3)}\!$}. Now, since the symbols $b_{11}$ and $b_{13}$ have been provided by {\small$\!{\rm Serv}^{(1)}\!$} and {\small$\!{\rm Serv}^{(3)}\!$} respectively, the symbol $b_{12}$ is recoverable because $(b_{11},b_{12},b_{13})$ is a codeword in the $[3,2]$ MDS code. Likewise, the symbols $b_{21},c_{12},c_{21}$ are also recoverable from the first two lines. Therefore, in the next two lines these recoverable symbols are combined with $a_{ij}$'s that have not been queried so far. Finally, the last two lines of Table \ref{fg1} are formed by applying (a1) and (a2) respectively.

The correctness condition holds because the user can finally obtain all $a_{ij}$'s for $1\leq i,j\leq 3$ and then recover $W_1$ by multiplying $S_1^{-1}$. To explain the privacy condition, we first point out a key observation of Table \ref{fg1}:
for $1\leq j\leq N$, each entry in the $j$th column of the matrices in (\ref{eq22}) appears in {\small$\!{\rm Serv}^{(j)}\!$}'s answers exactly once. Therefore, the answers of any two servers are  formed by symbols from two columns of the matrices which are completely independent because of the $[3,2]$ MDS encoding. Then plus the record symmetry enforced within each server, the privacy condition follows.

Finally, the total number of downloaded symbols from all servers is $7\!+\!6\!+\!6\!=\!19$ and each record consists of $9$ symbols. Hence the rate is $\frac{9}{19}$, which matches the capacity  for this case.

\section{Description of The General $T$-PIR Scheme}\label{sec3}

Now we describe how our $T$-PIR scheme works generally for $M\!\geq\!2$ and $1\!\leq\!T\!<N$. Suppose  {\small$W_1,...,W_M\in\mathbb{F}_q^{L}$}, and the user wants to privately retrieve {\small$W_{\theta}$} for some $\theta\in[M]$. Fix an $[N,T]$ MDS code over $\mathbb{F}_q$ with a $T\times N$ generator matrix $G$. It requires $q\geq N$ to ensure existence of such an MDS code. As in Section III, the general $T$-PIR scheme is described in two parts:

\vspace{0.1cm}

{\it Part I: Mixing-Rearranging.}

Let {\small$S_1,...,S_M$} be $M$ matrices chosen by the user independently and uniformly from all $L\times L$ invertible matrices over $\mathbb{F}_q$. As before, these {\small$S_i$}'s are only known by the user. We first assume $N|L$ and $L=N\tilde{L}$. Later we will show $L\!=\!Nn^{M\!-\!2}$ is enough for our scheme and this $L$ matches the optimal sub-packetization.

Define
{\small$
(U_{\theta})_{\tilde{L}\times N}=(u_{\theta,i}^{(j)})_{\tilde{L}\times N}={\rm Mat}_{\tilde{L}\times N}(W_\theta S_\theta),
$} where for the matrix entry $u_{\theta,i}^{(j)}$, $i$ is the row index and $j$ is the column index.
Then as in (\ref{eqb}),
define for $k\!\in\![M]\!-\!\{\theta\}$,
{\small\begin{equation*}
(U_{k})_{{\tilde{L}\!\times\! N}}\!=\!(u_{k,i}^{(j)})_{{\tilde{L}\!\times\! N}}\!=\!{\rm Mat}_{\tilde{L}\!\times \! N}\bigl(W_k S_k[:,(1\!:\!T\tilde{L})](I_{\tilde{L}}\otimes G)\bigr)\;.
\end{equation*}}

\vspace{-0.4cm}

\noindent Obviously, {\small$(U_{\theta})_{\tilde{L}\times N}$} is an invertible transformation of $W_\theta$, while {\small$(U_{k})_{\tilde{L}\times N}$} for $k\in\![M]\!-\!\{\theta\}$ is composed of independent rows each of which is a codeword in the $[N,T]$ MDS code. It is important to point out here the MDS encoding is used for achieving $T$-privacy and capacity of the PIR scheme. It has nothing to do with the database storage. In this work we consider replicated databases, not the MDS coded databases as in  \cite{Bana&Uluk16:CapacityPIRCoded,Wang&Sko16:CapaSymmePIR}.

\vspace{0.1cm}

{\it Part II: Combining}

In this part we build the answer table for general $T$-PIR schemes. Fix $k'\in\![M]\!-\!\{\theta\}$,  we first investigate how the entries of the matrix  {\small$(U_{k'})_{\tilde{L}\times N}$} are put into the answer table.

For any $\Lambda\subseteq[M]$ and $1\leq j\leq N$, we call {\small$\sum_{k\in\Lambda}{u}^{(j)}_{k,i_{k}}$} an $\Lambda$-type sum, where $i_{k}\in[\tilde{L}]$ is a row index. It is clear that each entry of {\small$(U_{k'})_{\tilde{L}\times N}$} appears in an $\Lambda$-type sum listed in the answer table for some $\Lambda\subseteq[M]$ with $k'\in\Lambda$. Particularly, define for all $k\in[M]-\{\theta\}$,

\vspace{-0.3cm}
$$\texttt{InType}_{k,\theta}=\{\Lambda\subset[M]\mid k\in\Lambda,\theta\not\in\Lambda\}\;,$$
\vspace{-0.5cm}

\noindent and for any $\Lambda\in\texttt{InType}_{k,\theta}$ denote $\underline{\Lambda}=\Lambda\cup\{\theta\}$.

First we associate each row of {\small$(U_{k'})_{\tilde{L}\times N}$} with a type $\Lambda\in\texttt{InType}_{k',\theta}$. Note that multiple rows may be associated with the same type. Moreover, because we always enforce record symmetry within each server, for any $\Lambda\in\texttt{InType}_{k',\theta}$ with fixed cardinality $|\Lambda|=i$, the number of rows associated with type $\Lambda$ remains the same and we denote this number as $d_i$. Moreover, the value of $d_i$ is independent of the record index $k'$. We will show how to determine $d_i$ for $1\leq i\leq M-1$ in Section \ref{sec4b}.
For convenience, we arrange these types in an increasing order of cardinality as they run through all sets in $\texttt{InType}_{k',\theta}$. As in the example, we have

\vspace{-0.4cm}
{\small\begin{equation}\label{eq33}{\scriptsize\begin{array}{cc} &\!\!\!{\rm type}\\\begin{pmatrix}
\!b_{11}\!&\!{\bf b_{12}}\!&\! b_{13}\!\\\!{\bf b_{21}}\!&\! b_{22}\!&\!b_{23}\!\\\! b_{31}\!&\!b_{32}\!&\!{\bf b_{33}}\!
\end{pmatrix}&\!\!\!\!\!\begin{array}{c}\{2\}\\\{2\}\\\{2,3\}\end{array}\end{array}\!\!,
~\begin{array}{cc} &\!\!\!{\rm type}\\\begin{pmatrix}
\!c_{11}\!&{\bf \!c_{12}}\!&\!c_{13}\!\\\!{\bf c_{21}}\!&\!c_{22}\!&\!c_{23}\!\\\!c_{31}\!&\!c_{32}\!&\!{\bf c_{33}}\!
\end{pmatrix}&\!\!\!\begin{array}{c}\{3\}\\\{3\}\\\{2,3\}\end{array}\end{array}}\end{equation}}
\vspace{-0.2cm}

\noindent which implies $d_1\!=\!2$ and $d_2\!=\!1$.

The entries of {\small$(U_{k'})_{\tilde{L}\times N}$} are put into the answer table according to the following rules (b1)-(b4):

\begin{itemize}
\item[(b1)] For $1\leq j\leq N$, entries in the $j$th column only appear in {\small${\rm Serv}^{(j)}$}'s answers.
\item[(b2)] For each row associated with a type $\Lambda\in\texttt{InType}_{k',\theta}$, $T$ out of the $N$ entries appear in $\Lambda$-type sums (i.e., interference symbols or interference sums) and the remaining $N-T$ entries appear in $\underline{\Lambda}$-type sums  (i.e., mixed sums).
\end{itemize}

For each $\Lambda\!\in\!\texttt{InType}_{k',\theta}$, denote by {\small$(U_{k'})_{\Lambda\times N}$} the matrix {\small$(U_{k'})_{\tilde{L}\times N}$} restricted to the rows associated with the type $\Lambda$.  Suppose $|\Lambda|=i$, then {\small$(U_{k'})_{\Lambda\times N}$} has $d_i$ rows.
\begin{itemize}
\item[(b3)] For each of the first $T$ columns of {\small$(U_{k'})_{\Lambda\times N}$}, $\alpha_i$ entries appear in $\Lambda$-type sums and the remaining $\alpha_{i+1}\!=\!d_i\!-\!\alpha_i$ entries appear in $\underline{\Lambda}$-type sums.
\item[(b4)] For each of the last $N-T$ columns of {\small$(U_{k'})_{\Lambda\times N}$}, $\beta_i$ entries appear in $\Lambda$-type sums and the remaining $\beta_{i+1}\!=\!d_i\!-\!\beta_i$ entries appear in $\underline{\Lambda}$-type sums.
\end{itemize}

One can check the scheme in Section III follows the above rules by combining (\ref{eq33}) with Table \ref{fg1}. Actually, in (\ref{eq33}) we have indexed in boldface the symbols which appear in mixed sums in Table \ref{fg1}. For example, in the first column of $(b_{ij})_{3\times3}$, $b_{11}$ appears in a $\{2\}$-type sum, which implies $\alpha_1=1, \alpha_2=1$, and $b_{31}$ appears in a $\{2,3\}$-type sum, which implies $\alpha_3=0$ and also $\alpha_2=1$. While in the third column of $(b_{ij})_{3\times3}$, both $b_{13}$ and $b_{23}$ appear in $\{2\}$-type sums, which implies $\beta_1=2, \beta_2=0$, and $b_{33}$ appears in an $\{1,2,3\}$-type sum, which implies $\beta_3=1$ and also $\beta_2=0$.

Actually, the rules (b1)-(b4) are due to our intention to guarantee correctness and privacy for the general scheme.

\subsection{Correctness and privacy}
Suppose $\Lambda\!=\!\{k_1,...,k_s\}\!\subseteq\![M]\!-\!\{\theta\}$. For $1\leq j\leq s$, the matrix {\small$(U_{k_j})_{\tilde{L}\times N}$} has $d_s$ rows associated with the type $\Lambda$, and all entries of the sub-matrix {\small$(U_{k_j})_{\Lambda\times N}$} appear in either $\Lambda$-type sums or in $\underline{\Lambda}$-type sums.
Moreover, we can assume locations of the entries that appear in
$\Lambda$-type sums are the same across all the sub-matrices {\small$(U_{k_j})_{\Lambda\times N}$}, $1\leq j\leq s$. For example, in (\ref{eq33}) we can see for $\Lambda=\{2,3\}$ the corresponding sub-matrices are

\vspace{-0.2cm}
{\small$$\begin{pmatrix}
b_{31}&b_{32}&{\bf b_{33}}
\end{pmatrix},~\begin{pmatrix}
c_{31}&c_{32}&{\bf c_{33}}
\end{pmatrix}.$$}

\vspace{-0.6cm}
\noindent Both have the first two entries as the ones to appear in $\{2,3\}$-type sums.

In general, for $1\leq t\leq d_s$ let $\Gamma_t\subset[N]$ be the set of column indexes that indicate locations of the entries in the $t$-th row of {\small$(U_{k_j})_{\Lambda\times N}$} to appear in $\Lambda$-type sums. Suppose the beginning row index of the sub-matrix {\small$(U_{k_j})_{\Lambda\times N}$} is $i_j$, $1\leq j\leq s$. Thus all the $\Lambda$-type sums in the answer table are

%\vspace{-0.1cm}
{\small\begin{equation}\label{eq44}\sum_{1\leq j\leq s}u_{k_j, i_j+t-1}^{(j')},~~j'\in\Gamma_t,~1\leq t\leq s\;. \end{equation}}

\vspace{-0.2cm}
\noindent
Accordingly, all the $\underline{\Lambda}$-type sums in the answer table are

%\vspace{-0.2cm}
{\small\begin{equation}\label{eq55}u_{\theta,*}^{(j'')}+\sum_{1\leq j\leq s}u_{k_j, i_j+t-1}^{(j'')},~~j''\in[N]-\Gamma_t,~1\leq t\leq s\;. \end{equation}}

\vspace{-0.2cm}
\noindent
Here we neglect the row index for the entry {\small$u_{\theta,*}^{(j'')}$} because each time we simply use a new desired symbol to form the mixed sums. Fix $t\in[s]$, from the MDS encoding we know {\small$$(\sum_{1\leq j\leq s}u_{k_j, i_j+t-1}^{(1)},\sum_{1\leq j\leq s}u_{k_j, i_j+t-1}^{(2)},...,\sum_{1\leq j\leq s}u_{k_j, i_j+t-1}^{(N)})$$}

\vspace{-0.2cm}
\noindent
is a codeword in the $[N,T]$ MDS code. Therefore, the interference parts in (\ref{eq55}) are recoverable from the $\Lambda$-type sums in (\ref{eq44}). As a result, all the desirable symbols involved in (\ref{eq55}), i.e., {\small$u_{\theta,*}^{(j'')}$}'s, can be obtained. As $\Lambda$ runs through all types in $[M]\!-\!\{\theta\}$, all desired symbols involved in the mixed sums can be obtained similarly, which give us a belief that the correctness condition can be satisfied. However, we still need to determine the exact number of desired symbols, which will be done in the next section.

About the privacy condition, rule (b1) first confines  {\small${\rm Serv}^{(j)}$}'s answers to the entries in the $j$-th columns of {\small$(U_k)_{\tilde{L}\times N}$}, $k\in[M]$, then rules (b3)-(b4) require each entry is involved exactly once in a symmetric form with respect to the records, i.e., for any $\Lambda\subseteq[M]$ with $|\Lambda|=i$, the number of $\Lambda$-type sums provided by {\small${\rm Serv}^{(j)}$} is $\alpha_i$ for $1\!\leq\!j\!\leq\!T$ and $\beta_i$ for $T\!<\!j\!\leq\! N$. Therefore, the answers of any $T$ servers  correspond to $T$ columns of {\small$(U_k)_{\tilde{L}\times N}$}, $k\in[M]$, combined in a symmetric manner. From the mixing-rearranging part we know any $T$ columns of the matrices are completely independent, therefore distribution of the answers of any $T$ servers is independent of the retrieved index $\theta$.

\subsection{Determining  $\alpha_i$ and $\beta_i$}\label{sec4b}
We derive necessary conditions on $\!\alpha_i\!$ and $\!\beta_i\!$ for feasibility of the rules (b1)-(b4) in the following:
{\small\begin{equation}\label{eqab}
  \left\{\begin{array}{l}
  T\alpha_i+(N-T)\beta_i=d_iT\\
  \alpha_i+\alpha_{i+1}=d_i=\beta_i+\beta_{i+1}\\
  \alpha_i,\beta_i\in\mathbb{N}, 1\leq i\leq M
  \end{array}
\right.
\end{equation}}
By a similar computation as that given in \cite{Zhang&Xu17:OptimalSubpacketization}, we obtain the solutions to (\ref{eqab}), i.e.,
for $N\geq 2T$,
{\small\begin{equation}\label{eqsolution1}
\renewcommand*{\arraystretch}{1.4}
\left\{\begin{array}{l}\alpha_i=\frac{(n-t)^{i-2}-(-t)^{i-2}}{n}(n-t)t^{M-i} \\
    \beta_i = \frac{(n-t)^{i-1}-(-t)^{i-1}}{n}t^{M-i}
\end{array}\right.\;,
\end{equation}}
and for $T<N<2T$,
{\small\begin{equation}\label{eqsolution2}
\renewcommand*{\arraystretch}{1.4}
\left\{\begin{array}{l}\alpha_i= \frac{t^{M-i}-(t-n)^{M-i}}{n}(n-t)^{i-1}\\
    \beta_i =\frac{t^{M-i-1}-(t-n)^{M-i-1}}{n}t(n-t)^{i-1}
\end{array}\right.\;,
\end{equation}}
where $n=N/d,t=T/d$ and $d={\rm gcd}(N,T)$. It is important to note that the values of $\alpha_i$ and $\beta_i$ are independent of the index $\theta$, which means the partial symmetric structure of the answer table is fixed for any given $N,T$ and $M$, no matter which record is to be retrieved. In other words, the partial symmetric structure can be publicly known, which will not affect the $T$-privacy.

Based on the values of $\alpha_i$ and $\beta_i$, we next compute sub-packetization of the scheme. For any $k\in[M]-\{\theta\}$, $d_i$ rows of {\small$(U_k)_{\tilde{L}\times N}$} are associated with a type $\Lambda\in\texttt{InType}_{k,\theta}$ with $|\Lambda|=i$, so $\tilde{L}\geq\sum_{i=1}^{M-1}\binom{M-2}{i-1}d_i$.
From (\ref{eqab}) we know {\small$d_i=\alpha_i+T\beta_i/(N-T)$}. From (\ref{eqsolution1}) and (\ref{eqsolution2}) one can compute that for $1\leq i\leq M$,
\begin{equation}\label{eqeq}
T\alpha_{i}+(N-T)\beta_{i}=d(n-t)^{i-1}t^{M-i}\;.
\end{equation}
Therefore, {\small$\tilde{L}\geq\!\sum_{i=1}^{M\!-\!1}\!\!\binom{M\!-\!2}{i\!-\!1}\!(n\!-\!t)^{i-1}t^{M\!-\!i\!-\!1}\!=\!n^{M\!-\!2}.$}
As for the matrix {\small$(U_\theta)_{\tilde{L}\times N}$}, its entries appear in $\Lambda$-type sums for all $\Lambda\subseteq[M]$ with $\theta\in\Lambda$. Particularly, for $1\!\leq\!j\!\leq\!T$, it needs $\tilde{L}\geq\sum_{i=1}^{M}\binom{M-1}{i-1}\alpha_i$ which implies $\tilde{L}\geq n^{M\!-\!2}$ from both (\ref{eqsolution1}) and (\ref{eqsolution2}), while for $T\!<\!j\!\leq\!N$, it needs $\tilde{L}\geq\sum_{i=1}^{M}\binom{M-1}{i-1}\beta_i$ which still implies $\tilde{L}\geq n^{M\!-\!2}$ from both (\ref{eqsolution1}) and (\ref{eqsolution2}).
Hence, it is enough to set {\small$\tilde{L}=n^{M\!-\!2}$} in our scheme. Thus the scheme has sub-packetization {\small$L=N\tilde{L}=dn^{M-1}$} which matches the optimal sub-packetization.

Then we compute rate of the scheme. It is easy to see the number of downloaded symbols is {\small$D= \sum^M_{i=1}\!\binom{M}{i}\bigl(T\alpha_i\!+\!(N\!-\!T)\beta_i\bigr)$} which equals $d(n^{M}\!-\!t^M)/(n\!-\!t)$ by (\ref{eqeq}).
Hence the rate of our scheme is $\frac{L}{D}=dn^{M-1}\!\times\!\frac{n-t}{d(n^M-t^M)}\!=\!\frac{1-T/N}{1-(T/N)^M}$, attaining the capacity for $T$-PIR schemes.

\subsection{Locator matrix}\label{sec4c}
In order to realize (b2)-(b4), we define a {\it locator matrix} $\mathcal{M}_{i}$ associated with each type $\Lambda\subseteq[M]-\{\theta\}$ with $|\Lambda|=i$ such that $\mathcal{M}_{i}$ is a $d_i\times N$ binary matrix whose $(i,j)$-entry equals $1$ if and only if the $(i,j)$-entries of the sub-matrices {\small$(U_k)_{\Lambda\times N}$} for all $k\in [M]-\{\theta\}$ appear in $\Lambda$-type sums. For example, for Table \ref{fg1} we have $\mathcal{M}_1={\footnotesize\begin{pmatrix}1&0&1\\0&1&1\end{pmatrix}}$ and $\mathcal{M}_2=\begin{pmatrix}1&1&0\end{pmatrix}$.  It follows immediately the $0$'s in $\mathcal{M}_{i}$ locate the interference parts of the $\underline{\Lambda}$-types sums. Specifically,
\begin{itemize}
  \item[(1)] by (b2) each {\bf row} of $\mathcal{M}_i$ has $T$ $1$'s, or, equivalently, has weight $T$.
  \item[(2)] by (b3) and (b4) each {\bf column} of $\mathcal{M}_i$ has weight $\alpha_i$ for the first $T$ columns and weight $\beta_i$ for the last $N-T$ columns.
\end{itemize}
Therefore, the general scheme can be implemented easily so long as we have built the binary matrices with the above two properties. We next give an explicit construction of these locator matrices.

Let $E(u,v)_{m\times n}$ denote an $m\!\times\!n$ binary matrix that has weight $u$ for each row and weight $v$ for each column. Evidently, a necessary condition for existence of $E(u,v)_{m\times n}$ is
\vspace{-0.3cm}
\begin{equation*}
\renewcommand*{\arraystretch}{0.85}
  \left\{\begin{array}{l}
  mu=nv\\0\leq u\leq n,~ 0\leq v\leq m
  \end{array}\right.\;.
\end{equation*}
\vspace{-0.3cm}

\noindent We can generally define $E(u,v)_{m\times n}$ as below:
\vspace{-0.2cm}
$$E(u,v)_{m\times n}={\footnotesize\begin{pmatrix}
{\bf e}_u\\{\bf e}_uP_u\\\vdots\\{\bf e}_uP_u^{m-1}
\end{pmatrix}}{~\rm where~}{\bf e}_u=(\stackrel{u}{\overbrace{1\cdots1}}~\stackrel{n-u}{\overbrace{0\cdots0}})$$
and $P_u\!=\!P^u\!,P\!={\scriptsize\begin{pmatrix}
{\bf 0}& { I_{n-1}}\\1&{\bf 0}
\end{pmatrix}}$. It is easy to see that ${\bf e}_uP_u$ is a cyclic right shift of ${\bf e}_u$ by $u$ coordinates.

Then combining with the solutions given in (\ref{eqsolution1}) and (\ref{eqsolution2}) one can verify the following matrices define the locator matrices needed for properties (1)-(2): for $N\geq 2T$,
$${\scriptsize\mathcal{M}_i=\!\renewcommand*{\arraystretch}{1.4}\begin{pmatrix}0&E(T,\beta_i)_{\frac{(\!N\!-\!T\!)\beta_i}{T}\!\times\!(\!N\!-\!T\!)}\\
E(T,\alpha_i)_{\alpha_i\times T}&0\end{pmatrix}}\;,$$
and for $T<N<2T$,
$$\mathcal{M}_i={\scriptsize\!\renewcommand*{\arraystretch}{1.4}\begin{pmatrix}E(2T-N,\frac{2T-N}{T}\beta_i)_{\beta_i\times T}&E(N\!-\!T,\beta_i)_{\beta_i\!\times\!(\!N\!-\!T\!)}\\E(T,\alpha_i\!-\!\frac{2T-N}{T}\beta_i)_{(\!\alpha_i\!-\!\frac{2T\!-\!N}{T}\beta_i\!)\!\times\!T}&0\end{pmatrix}}\;.$$

\subsection{Comparison with previous schemes}\label{sec3d}
In this section we make comparisons with previous capacity-achieving $T$-PIR schemes to give some insights into why our scheme can reduce both the sub-packetization and the field size. Since our scheme deals with PIR from $T$-colluding servers with replicated databases, we restrict to the comparisons with previous schemes under the same model,  such as the schemes in
\cite{Sun&Jafar16:ColludPIR,Zhang&Xu17:OptimalSubpacketization}.

First, reduction on the sub-packetization from the scheme in \cite{Sun&Jafar16:ColludPIR} is due to abandon of the symmetry across all servers. Instead, we divide all servers into two groups, the first $T$ servers in one group and the remaining $N\!-\!T$ servers in the other, and then enforce the symmetry across the servers within each group. That is, for any $\Lambda\subseteq[M]$ with $|\Lambda|=i$, the number of $\Lambda$-type sums provided by {\small${\rm Serv}^{(j)}$} is $\alpha_i$ for $1\!\leq\!j\!\leq\!T$ and $\beta_i$ for $T\!<\!j\!\leq\! N$. The same technique was used in \cite{Zhang&Xu17:OptimalSubpacketization} and the optimal sub-packetization is also achieved there. In one word, the asymmetry between servers means only a fraction of the answer table in  \cite{Sun&Jafar16:ColludPIR} is retained, however, that is enough for a capacity-achieving PIR scheme. On the other hand, the two-group partition and the partial symmetry within each group help to keep a neat implementation of the scheme.

Comparing with the scheme in \cite{Zhang&Xu17:OptimalSubpacketization}, we further reduce the field size to $q\!\geq\!N$, because in this work we replace the long MDS codes having information rate $\frac{T}{N}$ used in \cite{Zhang&Xu17:OptimalSubpacketization} with products of $[N,T]$ MDS codes. Therefore, an MDS code with much shorter length is sufficient here and the field size is reduced accordingly. Specifically, for any $\Lambda\subseteq[M]-\{\theta\}$ with $|\Lambda|=i$, the correctness condition in both work relies on the interference parts in all $\underline{\Lambda}$-type sums being recoverable as MDS parities from all $\Lambda$-type sums. However, in \cite{Zhang&Xu17:OptimalSubpacketization} the interferences are encoded as a whole by using an MDS code with rate $\frac{T}{N}$ and length $(T\alpha_i+(N-T)\beta_i)\frac{N}{T}=d_iN$, while in this work the interferences are encoded part by part through $d_i$ MDS codewords of length $N$. However, in order to maintain the privacy condition and the partial symmetric structure, we need to arrange these interferences in an properly designed order as elaborated by the locator matrices in Section \ref{sec4c}.

\section{Conclusion}\label{sec5}
The $T$-PIR scheme presented in this work deals with PIR problems from $T$-colluding servers with replicated databases. It simultaneously attains the highest rate and the smallest sub-packetization. Moreover, it only requires field size $q\geq N$ which is a great reduction  comparing with existing capacity-achieving $T$-PIR schemes.

\end{document}